# Molecular Control of Floquet Topological Phase in Non-adiabatic Thouless Pumping

Ruiyi Zhou[1] and Yosuke Kanai[1,2]

1. Department of Chemistry, University of North Carolina at Chapel Hill, Chapel Hill, North Carolina 27514, USA
2. Department of Physics and Astronomy, University of North Carolina at Chapel Hill, Chapel Hill, North Carolina 27514, USA

**Abstract**

Nonadiabatic Thouless pumping of electrons is studied in the framework of topological Floquet engineering, particularly focused on how changes to chemical moieties can control the emergence of the Floquet topological phase. We employ real-time time-dependent density functional theory to investigate the extent to which the topological invariant, the winding number, is impacted by molecular-level changes to *trans*-polyacetylene. In particular, several substitutions to *trans*-polyacetylene are studied to examine different effects on the electronic structure including mesomeric effect, inductive effect, and electron conjugation effect. Maximally-localized Wannier functions are employed to relate the winding number to the valence bond description by expressing the topological pumping as the transport dynamics of the localized Wannier functions. By further exploiting the gauge invariance of the quantum dynamics in terms of the minimal particle-hole excitations, the topological pumping of electrons can be also represented as a cyclic transition among the bonding and antibonding orbitals. Having connected the topological invariant to chemically intuitive concepts, we show the molecular-level control on the emergence of the Floquet topological phase, presenting us with a great opportunity for the intuitive engineering of molecular systems for such an exotic topological phase.

------------------

One of the earliest realizations that topological characteristics of Hamiltonian govern certain dynamical properties came from Thouless when he discussed the quantized pumping phenomenon in 1983[1]. Studying one-dimensional quantum-mechanical particle transport in a slow varying potential, he showed that the quantization of the number of transported quantum particles derives from the topology of the underlying Hamiltonian. Under the adiabatic evolution (i.e. instantaneous eigenstates of Hamiltonian), the particle current in a one-dimensional system is given by the topological quantity called winding number when the Hamiltonian is periodic in time. For topological materials, the winding number can take a non-zero integer value while it is zero for normal/trivial insulators. In recent years, Thouless pumping has been demonstrated experimentally in various systems[2-5] including an ultracold Fermi gas[6] and ultracold atoms in optical lattice[7]. Various theoretical studies[8-11] have employed model Hamiltonians such as the Rice-Mele model[12] and the description of topological pumping have often assumed a complete adiabaticity of the Hamiltonian evolution. At the same time, studies of non-adiabatic effects on Thouless pumping have begun to appear in the literature[13-15]. A non-adiabatic variance of Thouless pump has been proposed in a periodically-driven system or a Floquet system. The idea of the so-called topological Floquet engineering is to use a time-periodic field to induce topological properties in a driven system that is otherwise a trivial insulator[16-17]. In a Floquet system, the time-dependent Hamiltonian satisfies $\hat{H}(t+T) = \hat{H}(t)$ and time-independent effective Hamiltonian can be defined from the time evolution operator. One can analogously apply the topological description to this effective Hamiltonian. Under certain conditions, the Floquet topological phase, in which the winding number is a non-zero integer, can emerge. In our previous work using first-principles theory[18], we have shown that such a topological phase can be present for a *trans*-polyacetylene polymer chain using the external electric field as the driving field to obtain the Floquet condition. We have also shown that the Floquet topological phase can be directly linked to the well-established description of valance bond theory. There is an increasing effort to develop a molecular-level understanding of novel properties in topological materials from the perspective of chemistry[19-22], which would open up the field to be explored from a more intuitive viewpoint based on the arrangement of atoms. Having established the connection between the Floquet topological phase in the Thouless pump and the valance bond description, an important question now is to what extent this topological phenomenon can be controlled logically in terms of chemical understanding.

One can show the particle current in a one-dimensional (1D) system is given by the topological invariant called winding number when the Hamiltonian is time-periodic. For the topological phase, the winding number can take a non-zero integer value while it is zero for a normal/trivial insulator. This intricacy of Hamiltonian can be recovered conveniently from the phase information of quantum-mechanical wavefunctions (particularly important for studying real systems instead of model Hamiltonians), and it has a close connection to the modern theory of polarization developed in 90s[23]. In our recent work using first-principles theory[18], we demonstrated the non-adiabatic Thouless pumping of electrons in the Floquet topological phase in a trans-polyacetylene. In so-called topological Floquet engineering, a time-periodic field is used to induce a topological phase in the driven system that is otherwise a trivial insulator[16-17]. In a Floquet system, time-dependent Hamiltonian satisfies $\widehat{H}(t+T) = \widehat{H}(t)$ and time-independent effective Hamiltonian can be defined from the evolution operator over one time period $T$ such that $\widehat{H}_{eff}(k) \equiv iT^{-1} \ln \widehat{U}(k)$ where $\widehat{U}(k) \equiv \widehat{T} exp\left[-i \int_0^T dt\, \widehat{H}(k,t)\right]$ and $\widehat{T}$ is the time-ordering operator. Extending to the non-adiabatic regime, the winding number, being equal to the integrated particle current over the periodic time $T$, can be given in terms of the energy spectrum of the effective Floquet Hamiltonian, $\varepsilon_i$ (quasienergy)[24], or equivalently in terms of non-adiabatic Aharonov-Anandan geometric phase[25] of its eigenstates, $\Phi_i$. Conveniently for electronic structure theory, the winding number can be written as

$$W = \frac{1}{2\pi} \int_{-\pi}^{\pi} dk \sum_i^{Occ.} \langle \Phi_i(k, t=0) | \widehat{U}^\dagger(k) i \partial_k \widehat{U}(k) | \Phi_i(k, t=0) \rangle$$

$$= \frac{1}{2\pi} \sum_i^{Occ.} \left[ \int_{BZ} dk\, \langle u_i(k, t=T) | i\partial_k | u_i(k, t=T) \rangle - \int_{BZ} dk\, \langle u_i(k, t=0) | i\partial_k | u_i(k, t=0) \rangle \right] \quad (1)$$

where $u_i(k, t)$ are the time-dependent Bloch states in an extended 1D system. Note that the winding number is expressed analogously to the static Chern insulator, as $W = C \equiv \frac{1}{2\pi} \int_0^T dt \int_{BZ} dk \sum_i^{Occ.} F_i(k, t)$ where $C$ would the first Chern number, and the generalized Berry curvature is given by $F_i(k, t) = i[\langle \partial_t u_i(k,t) | \partial_k u_i(k,t) \rangle - \langle \partial_k u_i(k,t) | \partial_t u_i(k,t) \rangle]$ [26]. Due to the Blount identity $\langle w_i(t) | \hat{r} | w_i(t) \rangle = \frac{L}{2\pi} \int_{BZ} dk\, \langle u_i(k,t) | i\partial_k | u_i(k,t) \rangle$, the winding number can be expressed in terms of the time-dependent maximally-localized Wannier functions (MLWFs), $w_i(r, t)$, as[18]

$$W = L^{-1} \sum_i^{Occ.} [\langle w_i(t=T) | \hat{r} | w_i(t=T) \rangle - \langle w_i(t=0) | \hat{r} | w_i(t=0) \rangle] \quad (2)$$

where the position operator here is defined according to the formula given by Resta for extended periodic systems[27]. The winding number is then equal to the number of the geometric centers of the MLWFs (i.e. Wannier centers) pumped over the periodic time $T$ [24, 28]. This provides a real-space description of how the winding number physically represents the number of electrons pumped in time $T$. Kohn-Sham (KS) ansatz of Density Functional Theory (DFT) offers a convenient (and an exact, in principle) description for utilizing this single-particle theoretical formalism in studying real systems [29-36]. However, an additional mathematical complication exists because the KS Hamiltonian depends on the time-dependent electron density or the underlying KS orbitals. Using the KS ansatz for studying real systems, the Floquet condition (i.e. $\hat{H}_{KS}(t+T) = \hat{H}_{KS}(t)$) is therefore not necessarily satisfied even when the external driving field is time-periodic in $T$. This feature stems from the fact that time-dependent (TD) KS equations are non-linear differential equations, unlike TD Schrodinger equation. In our previous work[18], it was numerically shown that the time-integrated particle current $Q(T)$ indeed is an integer, being equal to the winding number, $W$, as expected when the Floquet condition is satisfied and Floquet topological phase (i.e. $W \neq 0$) can be found for trans-polyacetylene system. A brief discussion of our previous findings on unsubstituted trans-polyacetylene can be found in the Supporting Information.

An important question is to what extent the emergence of the Floquet topological phase, where the winding numbers are nonzero integers, is controlled by molecular-level changes in trans-polyacetylene. Answering this question at the atomistic level will be an important step toward developing the ability to rationally design materials with the Floquet topological phase. In this work, we consider four related systems as shown in Figure q. Fluorine substitution of C-H bonds in trans-polyacetylene (Fig. 1. (a)) is used to study the mesomeric effect; H atoms are substituted by electron-donating F atoms on one side of the carbon chain. The nitrogen substitution of C-C bonds (Fig. 1. (b)) is used for studying the inductive effect, which exhibits charge polarization of C-C $\sigma$ bonds. Having the higher electronegativity, one of the two distinct carbon atoms in individual monomer units are replaced with N atoms. We further consider the polyacetylene system with enol (OH) and enolate ion (O-) forms substituting a single C-H bond at a "defect" site as shown in Fig. 1. (c) and (d), respectively. The enolate ion case is particularly interesting because the valence bond model yields a well-recognized resonance structure with a C=O double bond, disrupting the electron conjugation on the carbon chain.

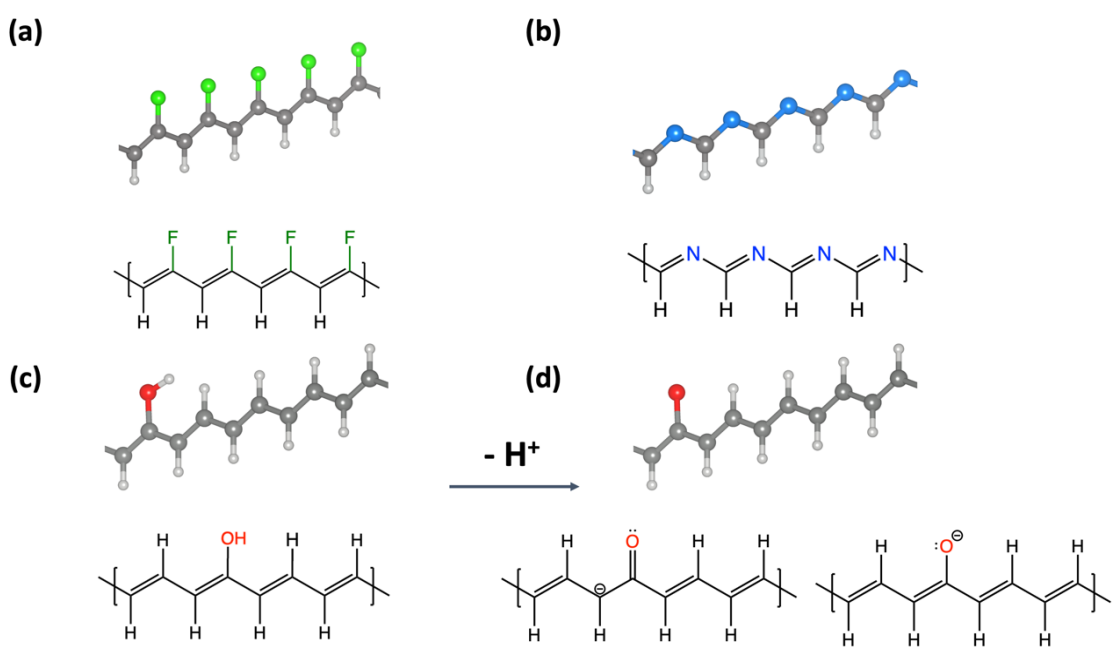

Figure 1. The chemical structures of four substituted trans-polyacetylene systems studied in this work: (a) Fluorine substitution at C-H bonds in trans-polyacetylene; (b) Nitrogen substitution at C-H bonds in trans-polyacetylene; (c) Enol (OH) substitution at a single C-H bond as "defect" site; (d) Enolate ion (O-) substitution at a single C-H bond as "defect" site.

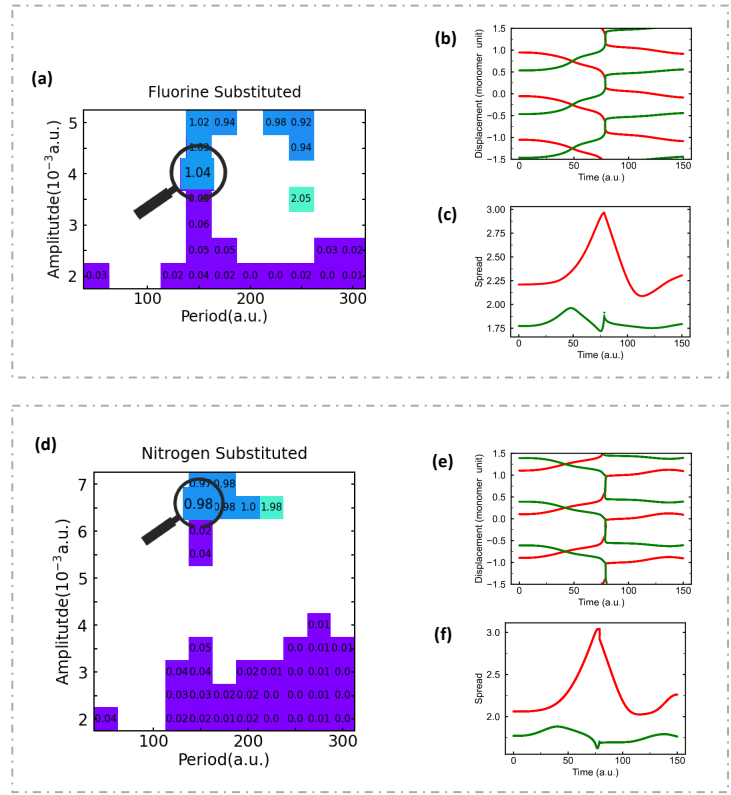

Figure 2. Time-integrated current Q(T) over one driving cycle for the fluorine-substituted (a) and nitrogen-substituted (d) trans-polyacetylene, plotted as a function of the electric field strength $|A|$ and the period $T$. The white areas represent that the Floquet condition is not satisfied. The red square box indicates the specific Floquet condition (W=1 topological phase) for which the MLWF dynamics were investigated in detail. For this specific case, MLWF dynamics are shown for the fluorine-substituted and nitrogen-substituted trans-polyacetylene: (b) spread values for the fluorine-substitution, (c) spread values for the fluorine-substitution, (e) displacement of MLWF centers for the nitrogen-substitution, and (f) displacement of MLWF centers for the nitrogen-substitution with C=C/C=N bonds (red) and C-C/C-N bonds (green).

Figure 2 (a) shows the calculated (time-) integrated current, $Q(T)$, per monomer for the field amplitude/period combinations for which the Floquet condition is satisfied for the F-substituted case. As can be seen, $Q(T)$ is quantized under these conditions as expected, and it is equal to the winding number (i.e. $Q(T) = W$) within the numerical accuracy anticipated from the RT-TDDFT simulation. For those cases in which the driving field amplitude is small, the winding number is zero and normal/trivial phase is present. For other areas, the Floquet topological phase can be identified such that the winding number is a nonzero integer. Although the exact combinations of the field amplitude and period for observing the Floquet topological phase differ from those of the unsubstituted trans-polyacetylene (Fig. 1), qualitative features appear rather similar. This is perhaps expected since the F-substitution does not significantly disrupt the electron conjugation that was found important for the topological pumping[18]. At the same time, the dynamics of the MLWFs show discernable differences. Let us focus on the specific case with the driving field of $T$=150 a.u. and $|A|$ =0.0040 a.u. for which the winding number is one. Fig. 2(b) shows that Wannier centers are transported much more abruptly when C-C and C=C Wannier centers cross among them (at $t \sim 1/2T$) as in a quantum tunneling-like behavior observed in Ref.[37], and their MLWF spreads become somewhat larger to the values of 1.92 and 2.96 $a.u.^2$, respectively (see Fig. 2(c)). The dynamical transition orbital (DTO) gauge is used for obtaining the minimal particle-hole excitation description of the quantum dynamics, and it was found to provide a convenient understanding of the topological pumping in Ref.[18]. In particular, a single DTO orbital, $\varphi_1^{DTO}(\mathbf{r},t)$, is largely responsible for the pumping behavior in this DTO gauge. Figure 3 shows how this particular DTO orbital transforms from having the π bonding orbital character at equilibrium to acquiring resonance and the π* antibonding characters in the driving cycle. Overall, the DTO transformation for the F-substituted chain is similar to that of the unsubstituted case, but with one notable difference. Due to the mesomeric effect, lone pair electrons on the F atoms shift the π electron density distribution, resulting in a noticeable delocalization of the DTO orbital onto the F atoms. However, the F-substitution does not severely disrupt the electron conjugation that is important for topological pumping, and the DTO changes remain largely the same as in the unsubstituted case.

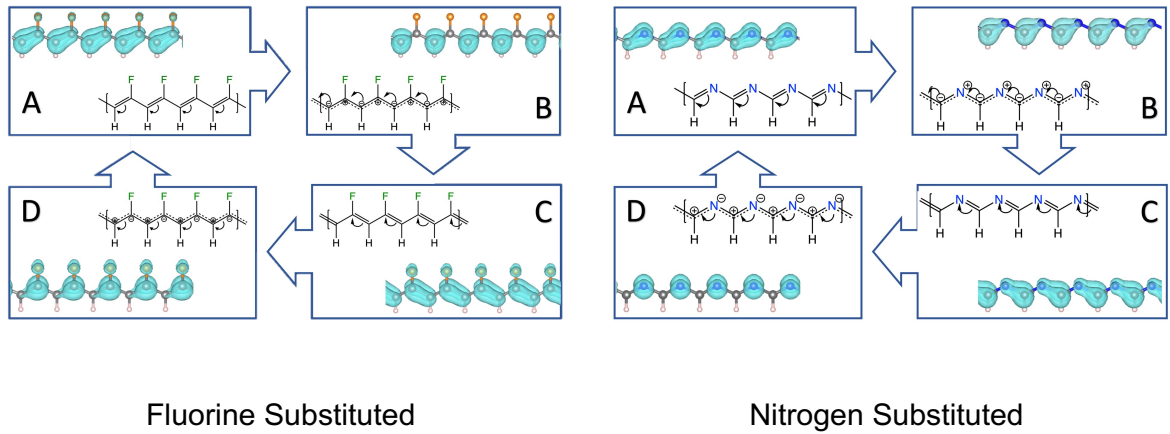

Fluorine Substituted            Nitrogen Substituted

Figure 3. Selected snapshots of the most dominant Dynamical Transition orbital (DTO) in one driving cycle for the fluorine substitution and the nitrogen substitution cases, with corresponding Lewis structures illustrating the electronic structure changes for the W=1 Floquet topological phase with $T$ =150 a.u./|$A$| =0.00400 a.u. and $T$ =150 a.u./|$A$| =0.00650 a.u. as marked in Fig. 2(a) and Fig. 2(d), respectively.

Figure 2 (d) shows the integrated current, $Q(T)$, for the nitrogen-substituted case. Just as in the case of the F-substitution of H atoms, the N-substitution of C atoms in the polyacetylene retains the existence of the Floquet topological phase. The valence bond model argument would suggest that the electron conjugation is preserved along the C/N chain, and the MLWFs indeed show the alternating arrangement for the double/single C-N bonds. At the same time, the C-N bonds are significantly polarized due to the high electronegativity of nitrogen atoms. The Floquet topological phase is observed in this case but for the driving fields that are quite different from those for the unsubstituted trans-polyacetylene case. In particular, a higher amplitude (the minimum of |$A$|=0.0065 a.u.) for the driving field was needed to induce the $W$=1 Floquet topological phase. This is likely due to the highly polarized nature of the C-N chain. Nitrogen atoms are more electronegative than carbon atoms, and a stronger perturbation is likely needed for the pumping of electrons than for the unsubstituted C-C chain case. Indeed, the DTO shows noticeably enhanced localization on nitrogen atoms than on carbon atoms as shown in Fig. 3 at $t$=0 a.u. when no external field is applied. Focusing on the specific case with the driving field of $T$=150 a.u./|$A$| =0.0065 a.u. ($W$=1), the Wannier center dynamics as shown in Fig. 2(e) show features that are like the F-substitution case. The tunneling-like behavior was again observed here, and the MLWF spreads for the C-C and C=C bonds become somewhat larger to the values of 1.67 and 3.04 $a.u.^2$, respectively (Fig. 2 (f)).

The tautomerism between -O⁻ and -OH substitutions for a C-H bond (see Fig. 1) represents a particularly interesting example for examining the relationship between electron conjugation and

Thouless topological pumping. Structurally, these two cases are quite similar and related only by a difference of a single proton. Figure 4 (a) and (b) show the integrated current, $Q(T)$, for these two cases. Even though these two structures are closely related by a simple proton, only the enol (-OH) case yields the Floquet topological phase where the winding number, $W$, is a nonzero integer while the enolate (-O⁻) case yields only the trivial insulator phase (i.e. $W = 0$).

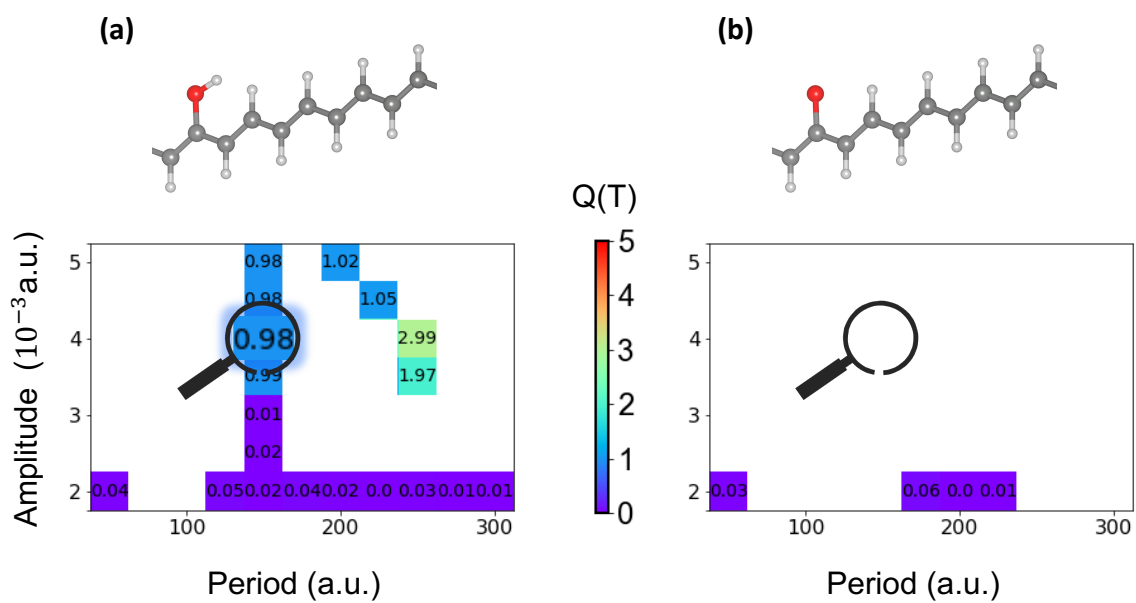

Figure 4. Time-integrated current Q(T) over one driving cycle for (a) the enol and (b) the enolate ion-substituted trans-polyacetylene, plotted as a function of the electric field strength |$A$| and the period T. The white areas represent that the Floquet condition is not satisfied. The marked region indicates the specific condition for which the MLWF dynamics were investigated in detail ($T$ =150 a.u. and |$A$| =0.00400 a.u.) as discussed in the main text.

While these two cases are structurally very similar, their valence bond structures are quite distinct. When a single enolate ion (-O⁻) is present, multiple resonance structures can be drawn by invoking a simple valence bond theory consideration as shown in Fig. 1 (d). In particular, one might easily depict a C=O double bond, and such a valence bond structure would result in the formation of a domain wall, disrupting the electron conjugation along the carbon chain. For relating the existence/absence of the Floquet topological phase to such a simple valence bond description, let us first discuss the MLWFs in the equilibrium state without any driving field applied. The enolate ion (-O⁻) structure contains three lone pair MLWFs associated with the oxygen atom while the enol (-OH) structure contains two oxygen lone pair MLWFs and one O-H bond MLWF. Table 1 shows that the MLWF spread of the C-O bond does not change significantly (only by ~0.02 $a.u.^2$) when a proton is removed from the enol (-OH) structure. However, the other three MLWFs localized on the oxygen atom become highly itinerant. Figure S6 in the Supporting Information shows that each MLWF in the enolate ion structure has a similar shape to those in the enol form. However, when compared with the enol form, the three lone pair MLWFs in the enolate ion are spatially more delocalized (i.e. the spread values increase by approximately 0.83 $a.u.^2$) with a significant contribution from adjacent carbon atoms. They indicate that parts of the lone pair electrons on the oxygen atom, for the enolate ion, are indeed involved in the conjugated carbon bonds through resonance, which is consistent with the resonance structures (Fig. 1(d)). At the same time, the C-C/C=C MLWFs in the enol (-OH) form are the same as the unsubstituted trans-polyacetylene case, with essentially the same MLWF spread values. The spread values of these C-C/C=C MLWFs do not vary among the monomer units, irrespective of their locations from the -OH defect site. However, in the enolate ion (-O⁻) form, the spread values of these MLWFs vary significantly along the C-C chain, depending on their location from the -O⁻ defect site.

| MLWF | OH | O- |
| --- | --- | --- |
| Lone pair | 1.4829 / 1.4829 | 2.3021 / 2.3019 |
| O-H bond / Lone pair | 1.6133 | 1.7499 |
| C-O bond | 1.3409 | 1.3637 |
| C=C | 2.3048 (0.0328) | 2.4098 (0.2358) |
| C-C | 1.7701 (0.0067) | 1.7928 (0.0381) |

Table 1: Spread values ($a.u.^2$) of the MLWFs in the enolate ion and the enol substituted cases.

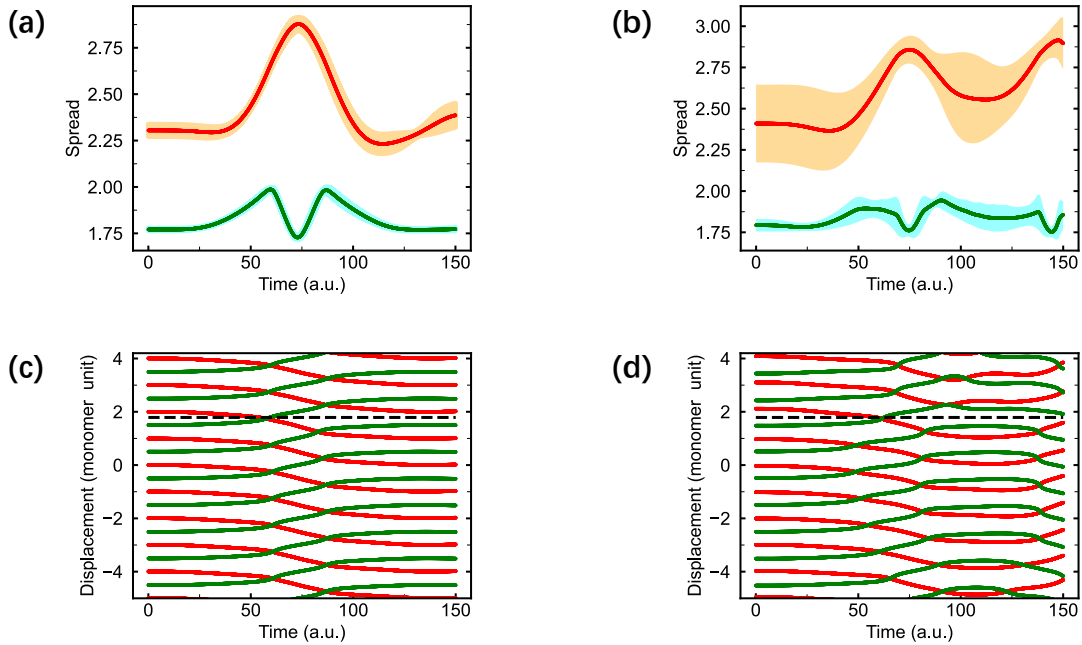

Figure 5. MLWF dynamics for the enol and enolate ion substitutions in trans-polyacetylene with the driving field of $T$=150 and $|A|$=0.00400 as marked in Fig. 4. (a) spread values for the enol substitution, (b) spread values for the enolate ion substitution, (c) displacement of MLWF centers for the enol substitution, and (d) displacement of MLWF centers for the enolate ion substitution. In (a) and (b), the average values are shown in the solid lines for C=C (red) and C-C (green) MLWFs while the shaded regions indicate the standard deviations among the MLWFs for each bond type. In (c) and (d), the red and green solid lines show the displacement of individual centers for C=C (red) and C-C (green) MLWFs and the dashed lines mark the positions of the OH/O⁻ defect sites on the chain.

Let us now examine the Wannier center dynamics to gain further insights into how the Floquet topological phase emerges in the enol (-OH) form while it is absent in the enolate ion (-O⁻) form despite their seemingly minor difference. The quantum transport characteristic is completely changed by the removal of a single proton. In particular, the driving field with $T$=150 a.u./$|A|$=0.00400, which yields a $W$=1 Floquet topological phase for the enol form while the Floquet condition is not satisfied for the enolate ion form, is discussed here. For the enol structure, the dynamics of MLWFs are essentially identical to that of the unsubstituted trans-polyacetylene, and the MLWF spreads do not vary much among them and also not in time (see Fig. 5 (a)). However, for the enolate ion structure, the MLWF spreads vary considerably in time, depending on their locations with respect to the -O⁻ defect site. The Wannier center dynamics for the enol and enolate structures are initially quite similar such that they move in the same direction continuously (see Fig. 5 (c) and (d)). However, at $t$ =100~150 a.u., C=C and C-C Wannier centers begin to reverse their movement in the opposite direction for the enolate ion (-O⁻) structure. The rates with which the C=C and C-C Wannier center change their moving directions depend on the distances from the -O⁻ defect site. The -O⁻ defect forms a domain wall such that MLWFs are prevented from transporting across the defect site to the C atoms on the opposite side.

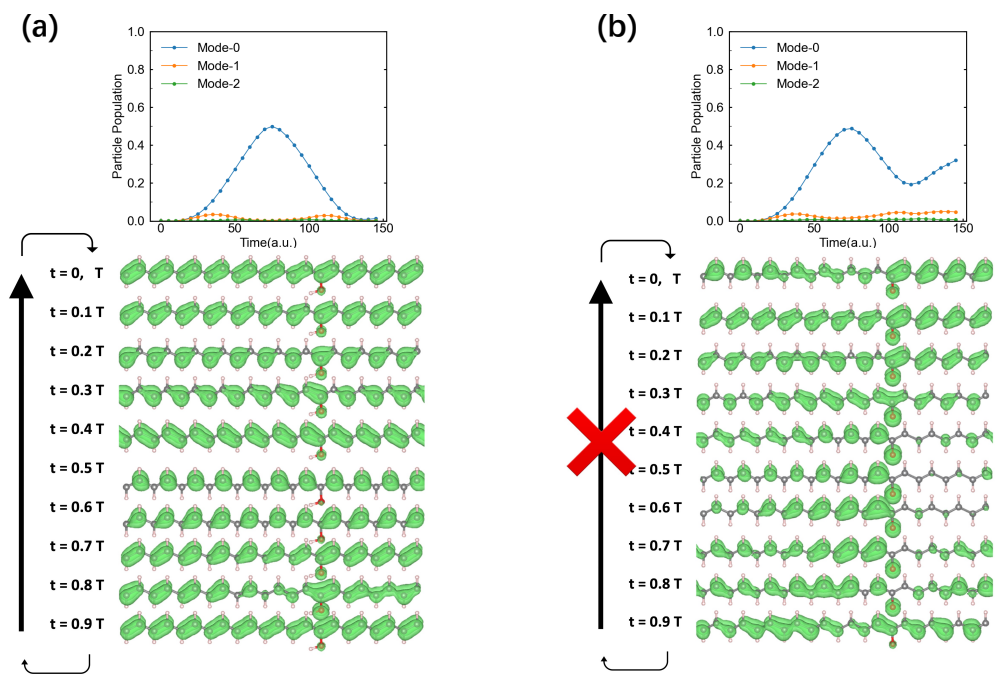

Figure 6. (Top) Time-dependent particle population for the three most dominant DTOs in terms of contributing to the dynamics and (Bottom) Snapshots of the most dominant DTO in a single driving cycle with the driving field of $T$ =150 a.u. and $|A|$ =0.00400 a.u. as marked in Fig. 4, for (a) the enol and (b) the enolate ion substitutions.

The quantum dynamics of electrons for both the enolate ion and enol structures still can be cast largely into the dynamics of a single DTO orbital. For the enol (-OH) structure, the particle population of this DTO orbital decays back to zero as shown in Fig. 6 (a), as perhaps expected from having the Floquet condition satisfied. For the case of enolate ion structure, the population does not decay back to zero even though this particular DTO is largely responsible for the whole quantum dynamics of electrons as seen in Fig. 6 (b). Snapshots of the DTO orbital transformation for both structures are also shown in Figure 6. For the enol structure, the introduction of the -OH substituting group has little effect on the topological pumping (see Fig. 6 (a)). The DTO transforms shows the changes from having the $\pi$ bonding orbital character at equilibrium to acquiring resonance and to acquiring the $\pi^*$ antibonding characters in a single driving cycle, exactly like for the unsubstituted case. However, Fig. 6 (b) shows that the DTO transformation is quite different in the case of the enolate ion structure for which the Floquet topological phase is not present. Already at t=0 when there is no driving field, the DTO appears substantially changed from the unsubstituted case (Fig S2 in Supporting Information). The DTO transformation is such that it is not possible to simply interpret its transition as the particle-hole transition between the $\pi$ and $\pi^*$ orbitals. Indeed, for the enolate ion case, the DTO orbital does not return to its original form at the end of the driving cycle because the Floquet condition is not satisfied.

Nonadiabatic Thouless pumping of electrons was studied in the framework of topological Floquet engineering, particularly focused on how atomistic changes to chemical moieties of *trans*- polyacetylene impact the emergence of the Floquet topological phase. In particular, we considered several atomistic substitutions to examine different types of effects on the electronic structure including mesomeric effect, inductive effect, and electron conjugation effect, based on our earlier analysis of the topological phase using the valence bond model[18]. For the cases with the substitution of hydrogen atoms with fluorine atoms (mesomeric effect) and the substitution of carbon atoms with nitrogen atoms (inductive effect), the Floquet topological phase is still present even though the specific driving field necessary for inducing the phase is different. Interestingly, the seemingly innocuous conversion of the C-OH substitution at a C-H site to its enolate ion form (C-O$^-$) led to the disappearance of the Floquet topological phase. The finding was rationalized by invoking the valence bond theory description, as the enolate ion form yields a prominent resonance structure featuring a C=O bond. The disruption of the electronic conjugation along the carbon chain by such a C=O resonance structure causes the Floquet topological phase to vanish. In summary, by connecting the topological invariant to the chemically-intuitive picture of the valence bond theory, we discussed how atomistic changes to trans-polyacetylene impact the emergence of the Floquet topological phase. The molecular-level understanding from the first-principles simulation enables a systematic and intuitive design of chemical systems for such an exotic topological phase. As in most theoretical works on topological materials currently, the atomic dynamics on the electronic Hamiltonian were not studied in this work. Our future work will examine the influence of the electronic current on the atomic lattice dynamics and reciprocally their impact on the non-adiabatic Thouless pumping of electrons as well as thermal effects on the atomic lattice.

## Computational Methods

Real-time time-dependent density functional theory (RT-TDDFT)[38-39] simulations were performed using the Qb@ll branch[31, 40-42] of the Qbox code[43] with a plane-wave pseudopotentials (PW-PP) formalism[44]. A 55-atom simulation supercell, consisting of 11 C-C monomer units aligned along the x-axis, was employed along with periodic boundary conditions (51.32 Bohr × 15.0 Bohr × 15.0 Bohr) and the Γ-point approximation for Brillouin zone integration was adopted. The molecular geometry of trans-polyacetylene (bond lengths, bond angles, and lattice constant) was taken to be that of the experiment[45]. The geometries for the substituted cases were generated by optimizing the atom positions of the substitution group while fixing other atoms. All atoms were represented by Hamann-Schluter-Chiang-Vanderbilt (HSCV) norm-conserving pseudopotentials[46-47], and the PBE[48] generalized gradient approximation exchange-correlation approximation was used with a 40 Ry plane-wave cutoff energy for the Kohn-Sham orbitals. For integrating the time-dependent Kohn-Sham equation, we employed the maximum localized Wannier functions (TD-MLWF) gauge and a time-dependent electric field was applied using the length gauge[37]. The enforced time-reversal symmetry (ETRS) integrator[49] was used with a 0.1 a.u. integration step size. Simulations were carried out with an applied electric field $E(t) = A \sin\left(\frac{2\pi}{T} t\right)$ as the driving field, and we considered the time period $T$ range of 50~300 a.u. and the field amplitude |$A$| range of 2.0~7.0 × 10$^{-3}$ a.u. with the uniform sampling intervals of 25 a.u. and 0.5 × 10$^{-3}$ a.u., respectively.

## Acknowledgment

This work was supported by the National Science Foundation under Award Nos. CHE-1954894 and OAC-17402204. We thank the Research Computing at the University of North Carolina at Chapel Hill for providing computational resources.

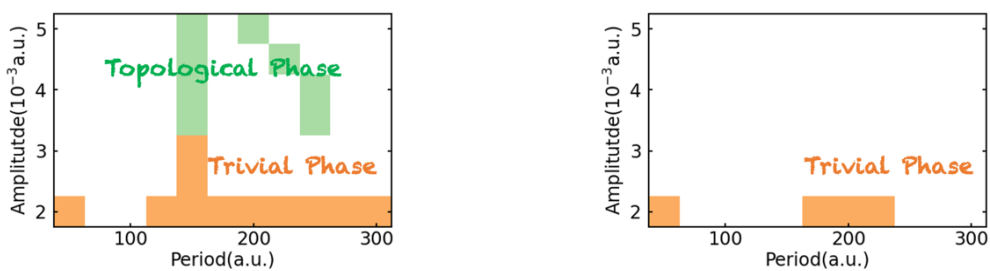

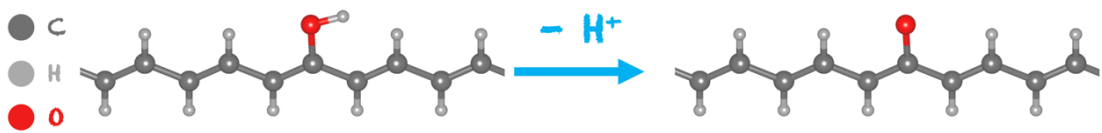

For Table of Contents Only